**Assessing the Effectiveness of Section 271 Five Years After the Telecommunications Act of 1996**

By Daniel R. Shiman and Jessica Rosenworcel[1]

October 2001

**I.     Introduction**

This paper examines the effectiveness of section 271 of the Telecommunications Act of 1996 (the Act) in achieving Congress' goals of increasing competition in the local and long distance telephone markets. In section 271, Congress developed an incentive structure whereby incumbent Bell Operating Companies (BOCs) are rewarded with entry into long distance markets in their territory if they can demonstrate to the Federal Communications Commission (FCC) that they have opened their local wireline markets to competition. In this paper we examine the logic behind this structure, to determine if it is a reasonable means of achieving increased competition in both the local and long distance markets.[2] We also provide an update on the extent of competitive entry in the local exchange market five years after enactment of the Act.

Some commentators have claimed that the section 271 process is unnecessary, alleging that it is a superfluous regulatory scheme that unnecessarily duplicates the market-opening provisions of section 251 (*e.g.*, Hausman and Sidak 1999, 430). There has also been extensive debate about the consequences of delayed BOC entry into the long distance market, with some economists arguing that immediate entry would provide the greatest benefit to consumers (*e.g.*, Crandall and Hazlett 2000, Crandall 1999). After analyzing the industry's technical and economic structure, and the legal and informational constraints under which regulators operate, we conclude that section 271 is a reasonably effective incentive mechanism for opening the local exchange market to competition, and is superior to other regulatory alternatives that are available. We also agree with those who argue that the use of entry into the long distance market as a prize for the BOCs will benefit the long distance market by adding an additional competitor only when that competitor's ability to vertically leverage its market power has been significantly reduced. Delaying entry until the local market is open to competition provides an appropriate safeguard against potential BOC discrimination against competitors in the long distance market (Schwartz 2000).

In section II we summarize the structure of sections 271 and 251 of the Act, and review the interrelation of section 271 and the market-opening provisions of section 251, in light of the history of the line-of-service restriction imposed by the Modification of Final Judgement. In section III we examine the economic, institutional and legal environment in which regulators operate, and discuss whether section 271 is likely to be an effective means of opening local telecommunications markets to competition, and whether BOC entry into the long distance market should be linked to local market access obligations. Section IV provides a statistical assessment of the development of local competition across the country, using available data, while section V offers some concluding remarks.

## II. Development of Section 271 of the Telecommunications Act of 1996

### A. The Local Exchange Market Prior to the Act

For over half a century AT&T's telecommunications network, which included both local and long-distance services, effectively was treated as a natural monopoly. In 1974 the Department of Justice initiated a lawsuit against AT&T under the Sherman Antitrust Act, alleging that AT&T used its local exchange monopoly to prevent competitive entry into the long distance and equipment manufacturing markets. The trial ended in a 1982 settlement known as the Modification of Final Judgment (MFJ). Under a key feature of the MFJ, AT&T divested itself of its local exchange telephone companies in 1984. These local exchange companies were organized into seven regional BOCs. The BOCs were subject to line-of-service restrictions that, among other things, prohibited them from competing in the long distance market.

The rationale underlying the divestiture was that AT&T would be unable to exercise monopoly control over the long distance market once it had lost control of the local exchange market. Moreover, without the line-of-service restriction prohibiting BOC entry into the long distance market, the BOCs would have the incentive to misallocate costs and subsidize competitive long distance services with their monopolized local exchange services, and degrade the quality of access received by interexchange

carriers (552 F.Supp 131, 162, 165, 170-75).

Under section VIII(C) of the MFJ, a BOC could petition for relief from the line-of-service restriction prohibiting BOC entry into the long distance market, if it could successfully demonstrate to the court "that there is no substantial possibility that it could use its monopoly power to impede competition in the market it seeks to enter." No BOC, however, successfully petitioned under the MFJ to provide long distance services.[3]

### B. The Interrelationship of Sections 251 & 271 of the Act

The Telecommunications Act of 1996 fundamentally altered the regulation of local telecommunications markets in the United States. It replaced the MFJ's restrictions on the BOCs with a new set of regulations,[4] and provided a means for the BOCs to enter the long-distance market. In the Act Congress also directed the FCC to remove the barriers that historically protected incumbent BOC local exchange carrier monopolies from competition and to develop regulatory policies that promote competition in local exchange markets. Two key elements of the Act supply the framework for opening up the BOCs' networks to competitors: section 251 and section 271. Both are briefly reviewed below.

#### 1. Section 251

Section 251 imposed a new set of requirements on local exchange companies (LECs), including the duty to provide for number portability, dialing parity, reciprocal compensation, resale of all services provided, and access to rights of way it owns (47 U.S.C. sect. 251(b)). LECs classified as incumbent local exchange carriers (ILECs), which are firms that have traditionally provided local exchange service in a local area and presumably have substantial market power in their local exchange market (47 U.S.C. sect. 251(h)), have an additional set of obligations to open up their networks for use by competitors.

Under section 251, ILECs are required to provide resold services to competitive local exchange carriers (CLECs) (47 U.S.C. sect. 251(c)). Thus, a CLEC can purchase the same services an ILEC makes available to its end-user customers at a discount that reflects the retail price minus avoided costs.

In addition, section 251 requires ILECs to provide access to the piece parts of their networks as separate, unbundled network elements (UNEs). In effect, this allows a CLEC to lease only those portions of the ILEC's network it wishes to use as inputs in providing its own telecommunications services to its own end-user customers. At present the FCC requires ILECs to provide UNE access to 7 UNEs, including loops (the transmission facilities between the demarcation point at the end-user's premises and the ILEC's central office), local circuit switching (equipment that determines the routing of local calls), and dedicated and shared transport (interoffice transmission facilities) (47 C.F.R. sect. 319).[5] These UNEs must be offered at cost-based prices, based on the pricing standards in section 252 of the Act, which the FCC has implemented as the marginal cost of offering the element in a hypothetical, technologically efficient network under the TELRIC (Total Element Long-Run Incremental Cost) standard.[6] CLECs may order these elements on an individual basis or as part of a complete package known as the UNE platform (UNE-P). In addition, an ILEC is required to permit interconnection at any technically feasible point, and provide for the collocation of equipment necessary for a competitor's interconnection or access to UNEs.

Finally, ILECs are required to negotiate in good faith interconnection agreements with competitors wishing to arrange interconnection and collocation, and purchase the ILECs' services through resale and purchase of UNEs. The process and deadlines for negotiating agreements, and for state arbitration of open issues, are laid out in section 252. State public utility commissions (PUCs) are given substantial responsibility for applying sections 251, 252, and the FCC's rules implementing those sections, to arrangements and interconnection agreements in their states.

Thus section 251 effectively provides for three modes of entry for CLECs: full facilities-based, partially facilities-based, and resale of the ILEC's services. Full facilities-based entry, in which the CLEC provides its own facilities and only needs interconnection and rights of way from the ILEC, has the potential to provide significant benefits to consumers, because CLECs are incented to develop new services, and bring new technologies to market, in order to differentiate their product offerings from the ILEC. Yet, scale economies outside of populated, downtown areas rich with business customers are unlikely to justify the cost of deploying new network facilities. The ability to enter using partial facilities-based entry allows for efficient choices regarding network duplication. Thus, CLECs may be able to rely on those UNEs, such as the local loop, where the economies of scale and scope are large and make entry difficult, and combine them with their own complementary facilities where deployment costs are justified by potential revenues, such as switching (an entry strategy known as UNE-Loop or UNE-L). This allows them the limited ability to provide their own set of unique services, without incurring the significant costs of duplicating bottleneck facilities.[7] Finally, entry through resale of the ILEC's services (which includes UNE-P) requires that the CLEC provide no transmission facilities of its own, allowing CLECs to reach customers for whom the cost of building facilities to serve them is too high to be practical, but limits them to providing the same services that are offered by the ILEC, over the same facilities.[8]

### 2. Section 271

If in section 251 Congress supplied the basic principles for local competition, by describing the terms and conditions for interconnection, resale, and access to unbundled network elements, then in section 271, Congress supplied the structured incentive for the largest ILECs, the BOCs, to abide by these principles and demonstrate their compliance.

Under section 271, a BOC must meet four requirements in order to receive authority from the FCC to provide in-region long distance services. Under the first of these four requirements, known as the Track A/Track B test, a BOC must show that either a facilities-based competitor currently exists within its market (Track A) or that the BOC has offered to provide competitors with access and interconnection to its network but no alternative provider has chosen to accept the offer (Track B).[9]

Under the second requirement, a BOC must demonstrate compliance with a 14-point competitive checklist. This checklist features a list of 14 discrete access requirements that reflect the basic interconnection principles of section 251. Among other items, the checklist requires that a BOC demonstrate that it provides interconnection, nondiscriminatory access to the UNEs required under section 251 at cost-based prices, local dialing parity, and the poles, ducts, conduits, and rights of way owned by the BOC. The checklist also requires that a BOC demonstrate that it makes its own retail services available for resale.

Under section 271's third requirement, a BOC must demonstrate how it will comply with section

272 of the Act, which requires that any BOC long distance services be provided through a separate affiliate, with separate books, officers and employees, and with all transactions between the BOC and the affiliate conducted at arm's length. These structural safeguards amount to a congressional acknowledgement that following section 271 approval, a BOC may still have an incentive to discriminate in providing exchange access services and facilities that rival interexchange carriers (IXCs) need to compete (FCC 1996b, 21911-13). Finally, the fourth requirement under section 271 requires that the FCC determine that a BOC's entry into the long distance market in a particular state is in the public interest (47 U.S.C. sect. 271(d)(3)(C)).

The FCC is required to make its determination on an application within 90 days of filing. Section 271 also specifically requires that the FCC consult with the Department of Justice and the state commission, and give "substantial weight" to the Department of Justice's evaluation of the BOC's filing.

As of this writing, the BOCs have filed a total of 20 section 271 applications with the FCC. During the first three years following enactment, the FCC denied 5 applications and 1 applicant chose to withdraw its filing. In 1999, the FCC approved its first section 271 application, for Bell Atlantic New York. Since that time, the FCC has approved 7 applications and 3 applicants have chosen to withdraw their filings. In addition, as of this writing, 4 applications are pending at the FCC. Table 1 provides a brief overview of the BOC applications filed since enactment of the Act.

**Table 1: Resolution of Past Section 271 Applications**

| Applicant | State | Filing Date | Resolution Date | Resolution |
|---|---|---|---|---|
| Ameritech | Michigan I | 1/2/97 | 2/11/97 | Withdrawn |
| SBC | Oklahoma I | 4/11/97 | 6/27/97 | Denied |
| Ameritech | Michigan II | 5/21/97 | 8/19/97 | Denied |
| BellSouth | South Carolina | 9/30/97 | 12/24/97 | Denied |
| BellSouth | Louisiana I | 11/6/97 | 2/4/98 | Denied |
| BellSouth | Louisiana II | 7/9/98 | 10/13/98 | Denied |
| Bell Atlantic | New York | 9/29/99 | 12/22/99 | Approved |
| SBC | Texas I | 1/10/00 | 4/5/00 | Withdrawn |
| SBC | Texas II | 4/5/00 | 6/30/00 | Approved |
| Verizon | Massachusetts I | 9/22/00 | 12/18/00 | Withdrawn |
| SBC | Kansas | 10/26/00 | 1/22/01 | Approved |
| SBC | Oklahoma II | 10/26/00 | 1/22/01 | Approved |
| Verizon | Massachusetts II | 1/16/01 | 4/16/01 | Approved |
| SBC | Missouri I | 4/4/01 | 6/7/01 | Withdrawn |
| Verizon | Connecticut | 4/23/01 | 7/20/01 | Approved |
| Verizon | Pennsylvania | 6/21/01 | 9/19/01 | Approved |
| SBC | Missouri II | 6/21/01 | 11/18/01 (Deadline) | Pending |
| SBC | Arkansas | 8/20/01 | 11/18/01 (Deadline) | Pending |
| BellSouth | Georgia | 10/2/01 | 12/31/01 (Deadline) | Pending |
| BellSouth | Louisiana III | 10/2/01 | 12/31/01 (Deadline) | Pending |

### III. Section 271 as an Appropriate Market-Opening Mechanism and Safeguard Against Discrimination

To determine whether the incentive structure of section 271 is an appropriate mechanism for increasing the competitiveness of local telecommunications markets, we examine two issues in this section of the paper: (1) whether section 271 is the best means of achieving the market-opening goals of section 251, given the informational and legal constraints regulators labor under; and (2) whether it is rational to link BOC entry into the long distance market to BOC success in opening the local market to competition.

#### A. Increasing Competition in the Local Market: Is Section 271 the Best Method of Achieving the Goals of Section 251?

##### 1. Background – The Challenge Facing Policymakers

This paper does not concern itself with the question of whether the goals of section 251 are economically desirable. Given the ILECs' enormous costs of opening up their networks to competition, the significant scale economies involved in providing local telecommunications services, and the regulatory costs involved in forcing reluctant incumbent providers to open their networks to competitors, it is at this point unclear whether the future benefits of increased competition will outweigh the costs of implementing section 251. For the sake of this paper, we assume that section 251's goals are achievable and economically desirable, and that the long-run expected net benefits to society and to consumers from implementing such a scheme is positive.[10] The legislator's/regulator's problem is to design an incentive mechanism to best achieve these goals.

There is a substantial theoretical and empirical literature on the design of incentive mechanisms for the communications industry (see, *e.g.*, Berg and Foreman 1996; Laffont & Tirole 2000). These mechanisms solve the basic principal-agent problem, in which the regulator is attempting through an appropriate monitoring and reward structure to elicit a desired kind of behavior from a regulatee. Crucial factors involved in determining which type of mechanism is most efficient is the information the regulator has about the company's costs and capabilities, the regulator's ability to monitor the actions of the company, and the tools the regulator has available to provide incentives to achieve the right behavior (Laffont & Tirole 1993, ch. 1). As we show later, the nature of the informational asymmetries must be carefully considered before deciding which mechanism will be most effective.

Two key well-known characteristics of the industry make opening up the local telecommunications market to competition difficult. First, the telecommunications industry is characterized by large network effects. The value of the network depends on the number of subscribers, making larger networks more valuable to join. Thus interconnection with the incumbent LEC is essential for new entrants and small carriers to be able to survive and compete. Second, local telecommunications networks have large fixed costs for certain parts of their operations, leading to significant economies of scale.[11] As a result, there are bottlenecks in the production of local exchange services for many kinds of customers. For instance, it is generally believed that for the majority of subscribers, the scale economies involved make duplicating the local loop by CLECs prohibitively expensive under current technologies.

Section 251 attempts to enable competition in the local exchange market in the face of these network effects and scale economies by requiring all ILECs to interconnect and to provide nondiscriminatory access to the piece parts of their networks as separate, unbundled network elements at cost-based prices, as well as provide their services for resale (47 U.S.C. sect. 251(c)).  By providing the means for three kinds of entry strategies (resale, full facilities-based, and partial facilities-based), section 251 increases the likelihood that competition will develop for all customers.[12]

Nonetheless, several factors make the implementation of the section 251 scheme difficult for regulators, particularly the technical complexity and closed architectures of the ILEC systems, and the lack of incentives for ILECs to cooperate.  The technical complexity of the ILECs' internal operations means regulators need a large amount of information to properly monitor and evaluate ILEC progress in making required changes to their systems.  The informational demands on regulators are further increased by the rapid rate of technological change, which forces frequent reassessment of the situation.[13]

Meanwhile, many critical aspects of ILEC networks, organizational structures, and back-end systems used to provision service are closed architectures.  By closed architectures, we mean systems designed to operate efficiently internally, but not designed for outside use or access to the intermediate stages of production and consumption.  Many of these architectures were developed prior to the Act, and therefore were constructed based on the assumption that services would only be used internally.  Consequently, to meet the requirements of section 251, ILECs are obligated to resolve a large number of technical and organizational problems, including: designing and implementing changes to their systems so CLECs can interconnect with their networks and utilize their network elements;[14] finding and constructing space for CLEC collocation of equipment; developing standardized interfaces and organizational procedures for CLECs to use to access their OSS computer systems to place orders and trouble tickets; generating technical documentation and building a support staff to support CLECs use of ILEC systems; creating new organizational structures to support the ILECs' wholesale operations, for example to manually process orders and handle troubles with CLEC lines; and working out new contractual and informational arrangements with CLECs.[15]  Thus, rather than view the Act's requirements as creating a static obligation for the ILECs to allow access to their networks by competitors, a more appropriate view might be to think of the Act as initiating a massive engineering project to convert the ILEC systems from closed to open architectures.[16]

Two important consequences follow from the technical complexity and closed architecture of the ILECs' networks. First, there is substantial uncertainty as to the time, effort and expense needed to complete the process of opening up the ILEC systems. Second, the regulator must possess an enormous amount of information to be able to assess whether the ILEC is implementing the processes appropriately. Thus it becomes difficult for a regulator, who often lacks technical expertise and first-hand knowledge, to determine whether the ILEC is making appropriate progress in making changes to its systems.

Usually regulators are able to rely on the regulated firm's natural incentives to provision a good.[17] ILECs, however, lack the incentive to open up their networks to their competitors. Not only is opening up their networks perceived as an additional and costly burden, but their success undermines the source of their market power (Economides 1999).[18] The development of a new contractual relationship between two firms normally requires significant effort and expense by the two firms to negotiate the terms and conditions of the contract, and to solve technical problems that may arise. Here ILECs potentially lack the incentive to negotiate in good faith, and to devote the physical and managerial resources needed to resolve the considerable number of problems that can be expected to arise from opening up their closed architecture network. When one side has a vested interest in preventing agreement, negotiations can bog down and be delayed indefinitely.[19] Regulators, therefore, must monitor the negotiation and interconnection process, but lack reliable tools for measuring good faith effort.[20] Marius Schwartz points out that it is easier for regulators to monitor existing arrangements than the negotiation and development of new ones (Schwartz 2000). The development of an appropriate monitoring and incentive mechanism thus becomes essential if regulators want to ensure that the ILECs' network is opened up to competitors in a timely manner.

### 2. Alternative Schemes

Congress had the freedom to choose which method it wanted to use to ensure that the BOCs, who were operating under the MFJ, would comply with section 251's requirements. When a new requirement is imposed on regulated companies, some thought must be given to developing a scheme to ensure efficient compliance. Taking into consideration the market characteristics just discussed, we next review a variety of alternatives to the section 271 incentive structure that might have been used to induce compliance with the market-opening provisions of section 251.

*Penalties for Observed Infractions.* One alternative incentive structure involves monitoring ILECs' behavior and imposing penalties for observed infractions. Kinds of penalties available include fines payable to the federal government, liquidated damages paid to affected CLECs, and suspension of an ILEC's participation in the in-region long distance market. Any fine or penalty used would need to be set large enough to deter noncompliance (Ford & Jackson 1999). This sort of scheme is the most common and most direct means of ensuring compliance with new laws and regulations. It is most effective when the requirements for compliance are clear, monitoring compliance is easy, and imposing fines or other enforcement penalties can be done directly and quickly.

For the purpose of implementing section 251, however, this method has significant drawbacks. The technical complexity of the network, and the substantial work needed to open up the ILEC's network and negotiate agreements for access and interconnection, makes it very difficult for a regulator to attempt to impose detailed requirements concerning what the ILEC must do, continually monitor the efforts made by the ILEC to comply, and spot and penalize every small violation. The regulator operates at a significant informational disadvantage, because of his lack of first-hand knowledge, resources, and equivalent technical expertise. The FCC's implementation of its Open Network Architecture (ONA) scheme—which predates the Act but similarly involves ILEC obligations to supply "unbundled" telecommunications inputs to third parties—illustrates how difficult it is for the regulator to resolve, as one commentator observed, "disputes over the details of what has—and has not—actually been implemented" (Schwartz 2000, 270).

In addition, the regulator has the burden of demonstrating that a fine or other enforcement action is warranted. The imposition of penalties for infractions will only be effective when both the standards for section 251 compliance and the evidence of malfeasance are clear. Because this kind of clarity is rare in this context, and is complicated further by political pressures and the threat of legal challenges, it would be difficult to initiate frequent, quick, decisive enforcement actions.

*Deadlines*. For regulators, outcomes are often easier to observe than the process. As a result, some of the disadvantages of the informational asymmetries facing regulators may be ameliorated if a deadline-oriented scheme is used. A deadline could be set by Congress, or the FCC could be allowed to set a deadline and/or grant extensions, with perhaps some penalty for noncompliance.

Determining how long a task should take, and setting the appropriate deadline, however, can be tricky, particularly if the task is technically complex. Regulators also may have a difficult time determining the merits of a regulated company's request for extension. For example, thirty-one carriers have asked for a waiver of the FCC's October 1, 2001 implementation deadline for wireless carriers to provide the locations of customers calling 911 (Stern, E1). The FCC has faced similar requests for an extension of the September 30, 2000 deadline for carriers to comply with packet-mode communications electronic surveillance requirements under the Communications Assistance for Law Enforcement Act (FCC 2001b). Regulators have the additional problem of deciding whether the company's actions are sufficient to be considered to have met the deadline. The pressure to pass may be considerable if an automatic penalty results from a declaration of noncompliance.

*Allow the States to Devise Their Own Schemes.* Alternatively, Congress could have encouraged the states to experiment with different regulatory structures in order to identify the market-opening schemes that generate the most benefits to consumers. Yet there are drawbacks to such a scheme. If each state has a different set of interconnection and unbundling rules, the ILECs would bear the significant costs of implementing a different scheme in each state. This scheme would also significantly raise the costs for CLECs attempting national or regional entry strategies, since they would have to develop different internal OSS systems to support a different scheme in each state, engage in more state specific negotiations with the ILEC, and lobby in each state to achieve the desired system. In addition, many state PUCs lack the technical personnel to design and implement an efficient scheme, and some state PUCs are more susceptible to lobbying from various parties, particularly the ILEC. Before passage of the Act only a few states had taken steps to open their local markets to competition.[21]

*Provide Direct Financial Compensation to the ILECs for Unbundling.* Congress and/or the FCC

alternatively could have provided a greater direct financial incentive for ILECs to interconnect and unbundle their networks. This could be achieved by setting the rates of unbundling at the cost of interconnection and provisioning the element, plus the full opportunity cost to the ILEC of providing the element. This pricing scheme is known as the Efficient Component Pricing Rule (ECPR), and was proposed by a number of economists in order to induce efficient levels of investment by entrants (Laffont & Tirole 2000; Baumol & Sidak 1994). It provides an incentive to the ILEC to provision UNEs to competitors when they are more efficient in provisioning complementary elements. The FCC rejected this rule, however, partly on efficiency grounds, because "the ECPR does not provide any mechanism for moving prices towards competitive levels; it simply takes prices as given" and "application of ECPR would result in input prices that would be either higher or lower than those which would be generated in a competitive market and would not lead to efficient retail pricing" (FCC 1996a, 15859-60).[22]

*Deregulatory Contracting Approach*. Congress could have chosen a deregulatory approach to the local exchange market. Unbundled access to network elements, interconnection, and resale could all be available only as a matter of private contracts between the ILECs and new entrants. Although the simplicity of this approach and the corresponding decrease in regulatory costs has some appeal, ILECs lack the incentive to enter into contracts at prices that would permit competitors to compete. New Zealand adopted this scheme in 1987, and competition has failed to develop there (Economides 1999).

### 3. Section 271: A Prize for Compliance

In the scheme adopted by Congress, requirements for interconnection and unbundling for ILECs are set out in section 251, and a large "prize" is given to the BOCs for demonstrating compliance in section 271.[23] The prize, of course, is allowing BOC entry into the long distance market, a market the BOCs were prohibited from entering under the MFJ. The BOCs are likely to enjoy significant profits from entry into the long distance market, due to the scope economies involved in provisioning both local and long distance services, resulting in substantial savings in administrative, marketing and overhead costs in serving long distance customers relative to the IXCs, and the low network costs involved, estimated at about one cent per minute (Schwartz 2000; Crandall & Waverman 1995; MacAvoy 1998).[24]

The use of a prize as a tool for ensuring compliance has several advantages over other incentive mechanisms available, especially penalties based on observed infractions and deadlines. First, it reduces the need for the regulator to monitor the pace of changes, or determine appropriate deadlines. Assuming clear and observable outcomes can be determined in advance, the BOC will have an incentive to meet those objectives, and therefore the need for the regulator to monitor progress, and determine if any delays are unwarranted, is greatly reduced.

Second, the regulator has an advantage legally in terms of the burden of proof. When awarding a prize, the burden of proof rests on the applicant to show that it has complied with the conditions of the prize. The imposition of fines or other enforcement penalties, on the other hand, requires the regulator to demonstrate that the penalty is warranted. While equal-sized penalties and rewards should have a symmetric impact on a firm's incentive to comply, the legal system's asymmetric treatment of penalties and rewards makes the use of rewards less cumbersome and potentially more effective. This is particularly the case with complicated and technically complex regulations, when the regulators' inherent incomplete information about the regulated firm makes it difficult to prove a violation has occurred. Use of a reward gives the company an incentive to provide information to the regulator in order to demonstrate it has complied. It also gives the regulator the flexibility to determine what proof the company needs to provide.

Third, utilizing a prize can be superior to threatening penalties if the legislator/regulator is fairly certain about the benefits of the new regulations, but is uncertain as to their cost, and therefore is unsure whether to mandate the change. If the total social cost from the change is equal to the company's private cost of implementing it, then by setting the prize equal to the total benefit to society from the new regulations, the company will have an incentive to implement the necessary changes only if the social benefit exceeds the social cost of making the change. In the case of section 271, however, the benefit of increased local competition from unbundling and interconnection is very likely to differ from the "prize" offered of entry into long distance, thus possibly leading to a sub-optimal outcome. The use of a direct monetary prize might have been more efficient, but is politically difficult to offer.[25]

Section 271 provides the prize of long distance entry if the BOCs demonstrate to the FCC that they have taken the necessary steps to open up their networks to interconnection with CLECs, and have implemented the processes necessary to provision UNEs to CLECs in a nondiscriminatory manner. Because it is difficult to observe a BOC's internal processes, the FCC has used the limited amount of freedom the Act gives it in choosing how to evaluate an application to urge that section 271 applicants submit evidence that make compliance more easily observable.[26] For instance, applicants rely on the submission of performance metric data, measuring the BOC's performance in providing service to CLECs, to demonstrate that they are provisioning services to the CLECs in a timely and non-discriminatory manner.[27] The FCC has also encouraged the use of independent third party tests to demonstrate that the BOC's OSS are readily accessible to the CLECs, for those areas of the BOC's operation difficult to measure with performance metrics (e.g., the quality of help documentation), and for areas where commercial volumes are too low to evaluate using performance metrics. Once 271 approval is granted the carrot for ensuring compliance is gone, but monitoring of post-approval compliance is made easier with the requirement that the BOCs continue to submit performance metric reports to the FCC.[28] The danger of post-approval backsliding has also been much reduced by the establishment of performance assurance plans that generate automatic payments by the BOC for poor performance to CLECs when indicated by the performance metrics.[29]

In addition, the section 271 process depends heavily on actions by state PUCs. Under sections 251 and 252 the state PUCs arbitrate interconnection agreements, and can also help develop a statement of generally available terms in order to expedite interconnection negotiations. The FCC has encouraged state PUCs to play an active role in the 271 process, by providing an independent evaluation of BOC compliance with the checklist, developing a third-party test of the BOC's OSS, creating performance metrics to measure BOC performance and provide for an independent audit of the BOC's data, and developing a performance assurance plan to prevent post-approval backsliding.[30] The BOCs have cooperated with state PUC efforts partly because the FCC's section 271 precedent relies heavily on the development of this kind of evidence, but also because state PUC endorsement of their section 271 application is helpful in seeking approval at the FCC. This has given the state PUCs some leverage to develop their own rules, and to develop a performance assurance plan.[31] Thus, the Act allows the state PUCs some room for experimentation, within the confines of the Act and FCC regulations. To a large extent, the success or failure of the Act to bring local competition to each state depends on the actions of the state PUC to arbitrate interconnection agreements, set wholesale prices, monitor BOC performance to CLECs, fix any deficiencies that appear in the performance metrics and the performance assurance plan, and handle complaints and take enforcement action if problems with BOC performance appear.

### B. Impact on the long distance market: Should BOC entry into long distance be linked to the BOC's implementation of Section 251?

While the use of a prize may be a superior solution to the problem of how to efficiently ensure BOC compliance with new interconnection and unbundling obligations, letting entry into long distance be the prize could prove costly if harm is done to the long distance market. Harm can occur from an unnecessary delay of entry into long distance, by eliminating an entrant capable of providing additional competition to the long distance marketplace, or from premature entry, allowing a BOC to vertically leverage its market power from the local to the long distance market. Some economists have argued that the long distance market is dominated by an oligopoly which has maintained prices above long-run incremental cost through tacit collusion, and suggest that delaying BOC entry to achieve some unrelated goal of opening up the local market denies consumers the benefit from adding a potentially powerful competitor to the long distance market (MacAvoy 1998; Crandall & Hazlett 2000).

There is, however, a long-standing concern on the part of regulators about the ability of a firm to vertically leverage its market power from one market to another through a vertical price squeeze or some other means. In the development of the MFJ, and in various proceedings of the FCC including the Competitive Carrier Proceedings, the ability of ILECs to vertically leverage their market power from the local market into other markets has been taken very seriously (FCC 1997). With a regulated monopoly in the local exchange market, BOCs will have an incentive to discriminate against rivals in the long distance market once they are granted entry in order to raise prices and reap the benefits of their market power (Schwartz 2000; Economides 1999).[32] BOC entry into long distance without adequate safeguards could cause harm to the long distance market and lower consumer welfare, if the BOCs are able to vertically leverage their market power into the long distance market.

BOC discrimination against rivals can happen in a variety of ways, including by footdragging in initiation of service or providing new features, by offering price structures that appear to be nondiscriminatory but in fact favor the BOC's affiliate, or by degrading the quality of a rival's service.[33] These forms of discrimination might be indistinguishable from legitimate conduct and difficult for regulators to detect or correct. Cooperation between interexchange carriers and their local exchange access supplier becomes more important over time as new technologies and the development of new call forward and distributing services make interconnection more complex and more difficult to arrange.

We won't get into a lengthy discussion in this paper of the long distance market and whether market prices reflect vigorous competition or collusive oligopoly. However, we believe that the potential for vertical leveraging of market power in the local exchange market into long distance is real, and that this threat outweighs any possible benefits from adding one more competitor to the long distance market. Therefore the preferred policy is to add a powerful entrant to that market, but only after ensuring that it is unable to vertically leverage its market power into the long distance market. This is best achieved by requiring the BOCs to open up their local market to competition before allowing them into the long distance market. If the BOC attempts to degrade the quality of interconnection with competitive IXCs, the IXCs' customers will have the opportunity to switch their local service to another local exchange provider that provides good quality interconnection services. Attempts to discriminate against competitive IXCs will then likely accelerate the BOC's loss of market share in the local market. Thus section 271 not only serves as an effective prize for opening up the local market, but also acts as a long-term guarantee that the BOC will lack the incentive to discriminate against its long distance rivals.

Section 271 just conditions BOC entry on the local market being open to competition, not on there being ubiquitous and pervasive competitive entry (such as would be required with a market share test). This can be justified by observing that it is the ability of consumers to switch local exchange companies that limits the BOCs' incentives to discriminate in the long distance market (Schwartz 2000). In addition, in many BOC markets CLEC entry may be slow to occur for a variety of reasons, such as a high cost of entry, low retail prices, and the level of demand in those markets. However, requiring the FCC and the state PUCs to certify that the market is open to competition places a heavy regulatory burden on these agencies. They must conduct detailed analyses to determine that the checklist has been met, and thereby demonstrate that there are no barriers to entry attributable to BOC failure to make the necessary changes to its systems and offerings or to state regulations. A market share test would have been much easier to apply, but would have delayed BOC entry for many markets that are less desirable to CLECs.[34]

## IV.    Statistics on the Development of Local Competition

While local competition has grown more slowly than many policymakers desired, there has been a steady increase in CLEC market share. According to data submitted in Form 477 to the FCC, and provided in the FCC's report "Local Telephone Competition: Status as of December 31, 2000," CLECs had 16.4 million lines for a market share of 8.5% of all switched end-user lines in the United States as of December 31, 2000. This was 97% higher than one year earlier. Data for total CLEC lines is unavailable for the period before December 1999, but data provided by the largest ILECs for lines provided to CLECs shows that the lines they have sold to CLECs, as resale and UNEs, have climbed rapidly, with an annual average growth rate of 62% for December 1997 to December 2000, as seen in Figure 1 (FCC 2001a).[35] Growth appears to have slowed recently, with the estimated number of loops provided by ILECs to CLECs for plain old telephone service (POTS) growing by 13% during the period from December 2000 to June 2001, for an annual growth rate of 28%.[36] This may be due to the well-documented woes of the CLEC industry, with many CLECs cutting back operations or filing for bankruptcy (Crandall 2001).

The level of competition across the country varies significantly, with CLECs gaining a market share as high as 20% in New York and 12% in Texas. The breakdown of CLEC market shares by state is provided in Figure 2, for those states with publicly available data (FCC 2001a).[37] States for which the FCC has approved a BOC's 271 application are shaded.[38]

While we have only limited publicly available information about the total number and location of CLEC lines in each state, we have better publicly available information about the number of lines purchased for POTS service from the BOCs and their affiliates, obtained from the performance metric reports that the BOCs file with state PUCs and with the FCC, for various purposes.[39] Using this data, we provide in Table 2 the breakdown by state of CLEC purchases of BOC services for providing POTS lines according to the type of facilities purchased, whether resale, UNE-P, or UNE-L for June 2001.[40] Figure 3 shows the breakdown by state for June 2001. While this table and figure does not include full facilities-based competition, and thus cannot be used to determine CLEC penetration, it does provide an indication of the extent of competition in each state.

The volume of lines provisioned to CLECs by ILECs can be considered a measure of how effectively ILECs have opened their systems and have set wholesale prices that allow for entry. It is generally more difficult for ILECs to meet their unbundling and resale obligations (because they may require major system changes) than it is for them to meet the needs of full facilities-based CLECs, which may only require basic interconnection similar to that already provided to IXCs. Consequently, this paper's data on lines purchased by CLECs may actually provide a better indication of how far the BOCs have progressed in meeting their Section 251 obligations than CLEC market share data. The percentage of lines provisioned to CLECs for resale, UNE-P and UNE-L for the larger ILECs for which data is publicly available is provided in Table 3, and shown in Figure 4 for December 2000. The much larger proportion of BOC lines ordered by CLECs in states with 271 approval (13.1% of BOC lines), or even in states that are eligible for 271 approval (4.2%), than for ILECs for whom section 271 does not apply (2.0% of ILEC lines), supports the argument that section 271 has been successful in providing an incentive for BOCs to open up their networks and allow for greater entry.[41]

The drawback to using this data is that the FCC's review of Section 271 applications, under the standard set by Congress, is not based on the extent of CLECs' penetration in the market, but instead whether the BOC has met the section 271 requirements, which effectively mean the market is open to competition. Thus, sections 251 and 271 do not guarantee that all forms of entry will be possible by efficient entrants, only that the cost of accessing the BOC's OSS systems, wholesale prices, the ability to get an interconnection agreement, and state and federal regulations will not be a hindrance to entry. Other factors have a major impact on the level of competitive entry in a state, including the level of retail prices, the quality and diversity of the ILEC's retail products, demand characteristics such as the wealth and density of customers and the number of businesses buying telecom services, the aggressiveness of the state commission in promoting competition, and the level of CLEC entrepreneurial activity in a state. For example, lower levels of CLEC entry into non-BOC ILECs' territories could be because smaller ILECs are usually located in rural areas, which may be less attractive markets to entrants.

From Figures 3 and 4 it is clear that in states with section 271 authority or with significant levels of CLEC purchases of lines, CLECs have tended to rely heavily on UNE-P as an entry strategy, especially in New York and Texas. In June 2001 UNE-P loops have been 65% of the lines purchased from BOCs in states with 271 approval versus 22% without it. When UNE-P prices provide a substantial discount from retail prices, which resale does not, UNE-P becomes an attractive means of reaching customers, especially price-sensitive residential customers who will only switch to achieve a significant savings.[42] Meanwhile, resale is a fairly popular entry strategy in most states.[43] In addition, while much has been made of CLECs' "cream-skimming" business customers while avoiding serving residential customers, data available suggests that CLECs are going after residential customers. Forty-one percent of CLEC lines in December 2000 were reported to be provided to residential and small business customers (FCC 2001a). In New York the state PUC reported that 52% of CLEC lines went to residential customers (New York Public Service Commission 2001).

We lack comparable consistent statistics measuring the growth of BOC entry on the long distance market, due to the recency of the BOCs' 271 approvals. However, Verizon reported in August 2001 it had signed up 6 million long distance customers, and had a market share of 31% in New York and 16% in Massachusetts (Verizon Press Release). SBC reported in July 2001 that it had gained 2.8 million customers in Texas, Kansas, and Oklahoma (SBC Press Release).[44]

V.     **Conclusion**

Section 271 has generated more controversy in the press and in political debate than it has in the economics literature. We hope that will change, because it represents a bold attempt to provide a strong incentive for ILECs to open up their networks, and cooperate with regulators, while providing protection for the long distance market. Sections 251 and 271 are part of an experiment to see if competition and rapid innovation can be brought to a market that has been considered a natural monopoly, and that historically has been slow to change in technology and products. There is much to learn from the successes and failures of this experiment, and future and foreign regulators could benefit from the insights gained by economists here.

The slow development of local competition has concerned many policymakers, but as section IV pointed out, there has been significant and continuous growth already. The process of opening up such a technologically complex industry with closed systems should have been expected to take many years. There were unrealistic expectations by many of the players (including CLECs, regulators, legislators, and investors) about how fast competition would develop. This process has been delayed further by litigation, conflicting court rulings, and disputes over who has the proper jurisdiction over each part of the process. A key part of the process is having all parties accept a common set of ground rules, and even now, this could be derailed by political pressures, new court challenges, regulatory fatigue, and the volatility of financial markets and the economy.

Based on our analysis of the industry's technical and economic structure, and the constraints under which regulators operate, we believe that section 271 is a more effective incentive structure for opening the local market to competition than other typically-used regulatory alternatives. We also believe that it is an appropriate safeguard for protecting competition in the long distance market after BOC entry has occurred. While full assessment will only be possible after twenty or thirty years, the statistical evidence we have five years after enactment of the Act suggests that the section 271 incentive mechanism has accelerated the process of opening local exchange markets to competition.

# Endnotes

[1] Daniel R. Shiman is an Economist and Jessica Rosenworcel an Attorney-Advisor in the Policy and Program Planning Division of the Common Carrier Bureau at the Federal Communications Commission. Opinions expressed are those of the authors alone, and do not represent the views or policies of the FCC or its commissioners.

[2] We do not address the appropriate level of prices associated with competitive entry in the local exchange market.

[3] Thus, the court generally concluded that the efficiency losses that would likely result from the BOCs' ability to engage in anti-competitive conduct outweighed the benefit of the BOCs' presence in the long distance market. *See generally BellSouth v. FCC*, 162 F.3d 678, 681 (D.C. Cir. 1998).

[4] The Act provided that "[a]ny conduct or activity that was, before the date of enactment [of the Act], subject to any restriction or obligation imposed by the [MFJ] shall, on and after such date, be subject to the restrictions and obligations imposed by the [Act] and shall not be subject to the restrictions and the obligations imposed by [the MFJ]" (Pub. L. No. 104-04, sect. 601(a)(1), 110 Stat. 56, 143).

[5] The other UNEs the FCC required to be unbundled are the Network Interface Device or NID, signalling and call-related databases, Operations Support Systems, and the high frequency portion of the loop (47 C.F.R. sect. 319).

[6] *See* 47 U.S.C. sect. 252(d); 47 C.F.R. sect. 51.501 *et seq.*; *Implementation of the Local Competition Provisions in the Telecommunications Act of 1996*, First Report and Order, 11 FCC Rcd 15499, 15844-47 (1996). Although the United States Court of Appeals for the Eighth Circuit stayed the FCC's pricing rules in 1996, the Supreme Court restored the FCC's pricing authority on January 25, 1999, and remanded to the Eighth Circuit for consideration of the merits of the challenged rules. *Iowa Utils. Bd. v. FCC*, 109 F.3d 418 (8$^{th}$ Cir. 1996), 120 F.3d 753, 800, 804-06 (8$^{th}$ Cir. 1997), *aff'd in part, rev'd in part sub nom.*, *AT&T Corp. v. Iowa Utils. Bd.*, 525 U.S. 366, 397 (1999). On remand from the Supreme Court, the Eighth Circuit concluded that certain FCC pricing rules are contrary to congressional intent. *Iowa Utils. Bd. v. FCC*, 219 F.3d 744 (8$^{th}$ Cir. 2000), *cert. granted sub nom. Verizon Communications, Inc. v. FCC*, 121 S.Ct. 877 (2001). The Eighth Circuit has stayed the issuance of its mandate pending review by the Supreme Court. *Iowa Utils. Bd. v. FCC*, No. 96-3321 *et al.* (8$^{th}$ Cir., Sept. 25, 2000).

[7] Many CLECs serving medium-sized business customers will purchase a high capacity (DS1 or higher) loop and transport to that customer and will aggregate the voice and data traffic onto that circuit.

[8] Resale is often used to serve business customers, especially those will to pay extra for better customer service or in locations that the CLEC's network does not yet reach, and residential customers who have failed to pay their bills to the ILEC.

[9] Thus, through the Track B test, a BOC operating in a state where there are no CLECs pursuing entry may nonetheless be eligible to apply for section 271 authority.

[10] It is possible that because of the ILEC's scale economies and low retail prices, and the presence of other barriers and difficulties hindering CLEC entry, that effective local competition could never develop. This paper assumes that effective local competition can be achieved through the right set of incentive and regulatory mechanisms.

[11] Because of these scale economies the local exchange market has been considered to be a natural monopoly, and regulated as such.

[12] Arguably, full-facilities based competition would yield the most benefits to consumers, not because it will lower prices the most, but because the potential long-term gains from innovation are greatest, including working around the bottleneck. However, CLECs may not find it economically feasible to use their own facilities to serve many kinds of customers.

[13] For example, technological changes may change the location of bottlenecks, may change the cost of unbundling particular elements, and may change the speed with which requirements may be implemented.

[14] Some parts of the network are difficult to unbundle for CLECs' use, such as digital loop carrier systems, which may have been employed in place of the traditional copper loop.

[15] This is only a partial list of the actions an ILEC must take to open its network to competitors' use. For a more detailed discussion, see FCC 1996a and the FCC's various orders in response to section 271 applications.

[16] The ambitious nature of the Act makes it highly regulatory in certain aspects, requiring detailed regulations governing the ILECs wholesaling of their networks and services, but with the goal of reducing regulation at the retail level. A similar purpose was served by Part 68 of the Code of Federal Regulations, which provides detailed rules governing the connection of customers premise equipment (CPE) to the network, in order to improve competition in the CPE market.

[17] Usually regulators are setting retail prices which the firm has an incentive to produce, with the price set above marginal cost. The Act appears to create a very unusual situation historically, in which firms are required to go to a great deal of effort to produce a good they do not want to sell.

[18] As has been widely recognized, there are circumstances in which upstream monopolists who are competing

downstream will not want to discriminate against users of their inputs, such as when they can extract their monopoly profits upstream, or when the downstream users are much more efficient, but those circumstances do not generally apply to ILECs (Laffont & Tirole 1999; Schwartz 2000). In addition, ILECs gain two benefits from degrading competitors' access, one from reducing their loss of retail customers, and the second from saving money by not developing the needed facilities for unbundling and interconnection.

[19] For a discussion of problems of this sort, see Morris and Preece 1982.

[20] It is very difficult to set performance standards and benchmarks concerning the negotiation of new arrangements and the working out of technical details. The Act provides for the use of arbitration and regulatory rule-making to resolve outstanding issues, but this process can be slow.

[21] New York and Illinois were early states to promote competition in the local exchange market (Tomlinson 2000, 169).

[22] ECPR was also rejected because it was inconsistent with the Act, since "the existing retail prices that would be used to compute incremental opportunity costs under ECPR are not cost-based" (FCC 1996a, 15859).

[23] The remainder of this section concentrates on the BOCs, since section 271 only pertains to them. They were the only ILECs prohibited from entering the long distance market under the MFJ, so Congress had the unique opportunity to use long distance entry as a prize for them. The BOCs are by far the largest ILECs, and have about 75% of the nation's access lines.

[24] Indeed, in states where it has been permitted to sell long distance service, SWBT only offers long distance service to customers that are also taking its local service.

[25] Tax credits, however, are sometimes used as a prize for undertaking some action such as an investment.

[26] The Act requires that the FCC process a lot of information about the BOCs' systems, since the requirement is only that the BOCs systems be open to competitors, and not that the market is competitive.

[27] "Non-discriminatory service" is interpreted to mean that service to CLEC customers is of the same quality and timeliness as service to the BOC's retail customers, when the same service is provided to both sets of customers.

[28] Under section 271(d)(6) the FCC has authority to take an enforcement action if the BOC's performance were to decline post-approval.

[29] This comports with Marius Schwartz's argument that the section 271 prize is needed so long as new arrangements are being made. Once the arrangements are established and reporting mechanisms in place, routine enforcement should be sufficient to prevent discrimination against CLECs (Schwartz 2000).

[30] The FCC has encouraged the use of open collaboratives, in which CLECs can participate and voice their concerns, in the state process. Where state PUCs lack the requisite resources to accomplish all of these tasks, they often rely on the efforts of large state PUCs in their BOC region to accomplish the more difficult region-wide tasks, such as conduct a third party test, or in some cases have pooled their resources with other state PUCs in the region.

[31] Many state PUCs lack the statutory authority to impose a plan that provides for automatic fines levied against the BOC, so with the FCC's encouragement, the BOCs have voluntarily submitted to a state-developed performance assurance plan.

[32] According to economic theory, a firm with an unregulated monopoly in one market will be unlikely to discriminate against rivals in a vertically-related market, since the firm should be able to extract all possible monopoly profits by raising prices in the monopolized market. However, there is an incentive to vertically leverage the market power if the monopolized market is regulated, such that the firm is unable to raise prices and extract monopoly profits from it.

[33] Degradation of a rival's service can occur by allowing interconnection to deteriorate, thus allowing more calls to be blocked to an IXC than calls to its own long distance affiliate. One BOC engineer has said that in the past it set up its tandems to have lower blocking (i.e., fewer calls unable to get through at peak calling times) for long distance calls than for local calls, since carrying long distance calls was much more profitable.

[34] A market share test would also have given CLECs owned by the IXCs the opportunity to game the process, by limiting their own entry into the local market to prevent BOC entry into long distance. The IXCs have been seen as natural entrants into the local market, and have aggressively entered many local exchange markets through a variety of entry strategies.

[35] We use the terms "purchased" and "sold" for convenience, although the lines are in fact being leased, if obtained as a UNE, or the services on that line are being resold by the CLEC.

[36] This estimate was based on numbers taken from the performance metric reports, discussed below. The number of UNE loops provisioned for DSL grew at about the same rate for this period.

[37] The data was withheld for about a third of the states for confidentiality reasons (FCC 2001a).

[38] The states counted as having section 271 approval for this section are those states currently having approval: New York, Texas, Kansas, Oklahoma, Massachusetts, Connecticut and Pennsylvania. Note that this graph includes all ILEC lines in the state, while 271 approval applies only to the BOC territory in these states. The data used comes from FCC 2001a, and is the number of switched access lines reported by each CLEC divided by the total number of switched access lines (CLEC plus ILEC) in the state. FCC approval is not based on market share lost to CLECs, which explains why some states with lower levels of CLEC entry have received section 271 approval.

[39] These reports show the level of performance by the BOC to the CLECs in a variety of areas and product lines, and compare it to performance to their own customers, where possible, or to a benchmark. The reports were developed for section 271 applications, and to meet the conditions of a merger approval. See http://www.fcc.gov/ccb/mcot/ for performance metric data filed for the SBC-Ameritech and Bell Atlantic-GTE mergers. The total number of CLEC lines for a particular product category for a state were taken from the December 2000 and June 2001 reported CLEC volumes (i.e., the denominator used) for the network trouble report rate metrics, which measure the percentage of all lines in a product category that had a reported trouble for that month. We consider this data to be more reliable than the form 477 data, since for both section 271 applications and for the merger conditions an audit of the data was required.

[40] We assumed that the number of ILEC lines remained the same as for December 2000 (according to FCC ARMIS data, growth of BOC access lines appears to have been about zero for 1999-2000). By the term "BOC" we are referring to states classified as in-region territories for section 271 purposes. Companies subsequently merged into the BOCs, such as SNET (now a part of SBC) and GTE (now a part of Verizon), are considered separately for the purposes of this discussion. Section 271's requirements do not apply for BOC affiliates in states the BOCs were not in when the Act was passed.

[41] The data for "BOCs in states with section 271 authority" are for the BOCs in all states that today have 271 authorization, for both December 2000 and June 2001. See fn. 38 for the list of those states.

[42] The long-term economic benefits from CLECs offering UNE-P and resale, however, are likely to be much less than those from CLECs investing in their own facilities, for full or partial facilities-based competition.

[43] Statistics available indicate that resale is used to serve business customers more – about 55% of resale POTS lines went to business customers in June 2001, from data for all of the BOCs except Verizon. In New York 90% of all resale lines went to business customers. Resale can be an easy and useful means of providing service to small and medium-sized businesses who are willing to pay extra for customized billing and additional customer care.

[44] This translates into about 43% of Verizon's own estimated 14 million retail lines in New York and Massachusetts and 27% of SBC's own 10.2 million retail customers in Texas, Kansas, and Oklahoma, excluding CLEC lines.

**Table 2: Percentage of BOC Switched Lines Purchased by CLECs, by Type of Line, for June 2001**

| State | BOC | Total BOC Lines | BOC % of ILEC Lines | Percent Resale | Percent UNE-P | Percent UNE-L | Percent Total | Total Lines |
|---|---|---|---|---|---|---|---|---|
| Alabama | BellSouth | 1,997,723 | 79% | 2.2% | 1.8% | 0.6% | 4.7% | 93,048 |
| Arizona | Qwest | 2,959,467 | 93% | 1.7% | 0.3% | 0.5% | 2.5% | 73,272 |
| Arkansas | SBC- SWBT | 1,048,587 | 69% | 3.3% | 0.5% | 1.6% | 5.4% | 56,340 |
| California | SBC- PacTel | 18,810,937 | 78% | 1.0% | 0.3% | 1.6% | 2.9% | 552,391 |
| Colorado | Qwest | 2,845,889 | 96% | 2.0% | 1.7% | 0.8% | 4.6% | 130,480 |
| Connecticut | Verizon- BA North | 57,893 | 99% | 3.5% | 0.0% | 0.9% | 4.4% | 2,547 |
| Delaware | Verizon- BA South | 595,708 | 100% | 1.8% | 0.0% | 2.9% | 4.7% | 28,061 |
| District of | Verizon- BA South | 1,019,026 | 100% | 7.8% | 0.0% | 0.9% | 8.7% | 89,023 |
| Florida | BellSouth | 6,850,656 | 59% | 3.0% | 1.4% | 1.6% | 6.0% | 408,159 |
| Georgia | BellSouth | 4,264,151 | 83% | 2.7% | 3.1% | 1.5% | 7.4% | 313,869 |
| Idaho | Qwest | 583,168 | 72% | 1.8% | 0.1% | 0.0% | 1.9% | 11,013 |
| Illinois | SBC- Ameritech | 6,880,696 | 85% | 3.9% | 1.9% | 3.0% | 8.8% | 607,957 |
| Indiana | SBC- Ameritech | 2,256,736 | 62% | 1.5% | 0.0% | 1.6% | 3.2% | 71,381 |
| Iowa | Qwest | 1,143,962 | 65% | 1.0% | 11.0% | 0.8% | 12.8% | 146,145 |
| Kansas | SBC- SWBT | 1,389,742 | 84% | 5.7% | 3.9% | 0.3% | 9.9% | 137,041 |
| Kentucky | BellSouth | 1,264,064 | 57% | 2.4% | 1.4% | 0.3% | 4.0% | 50,859 |
| Louisiana | BellSouth | 2,439,723 | 93% | 3.8% | 0.6% | 0.4% | 4.8% | 115,929 |
| Maine | Verizon- BA North | 749,853 | 83% | 5.3% | 0.2% | 1.2% | 6.8% | 50,868 |
| Maryland | Verizon- BA South | 4,051,759 | 100% | 5.5% | 0.1% | 1.0% | 6.6% | 267,007 |
| Massachusetts | Verizon- BA North | 4,636,622 | 100% | 5.8% | 0.6% | 1.8% | 8.2% | 378,294 |
| Michigan | SBC- Ameritech | 5,397,189 | 84% | 2.0% | 3.2% | 2.3% | 7.5% | 403,613 |
| Minnesota | Qwest | 2,342,669 | 73% | 3.1% | 3.4% | 2.1% | 8.7% | 203,010 |
| Mississippi | BellSouth | 1,359,773 | 93% | 4.4% | 1.0% | 0.3% | 5.7% | 77,687 |
| Missouri | SBC- SWBT | 2,605,726 | 75% | 4.1% | 2.3% | 0.3% | 6.7% | 173,384 |
| Montana | Qwest | 386,624 | 68% | 3.2% | 0.1% | 0.4% | 3.6% | 14,067 |
| Nebraska | Qwest | 507,263 | 50% | 2.2% | 0.4% | 1.3% | 3.9% | 20,035 |
| Nevada | SBC- PacTel | 380,616 | 27% | 2.1% | 0.0% | 1.6% | 3.8% | 14,285 |
| New Hampshire | Verizon- BA North | 816,322 | 94% | 5.5% | 0.3% | 2.9% | 8.8% | 71,801 |
| New Jersey | Verizon- BA South | 6,914,330 | 97% | 2.9% | 0.1% | 0.5% | 3.5% | 240,290 |
| New Mexico | Qwest | 863,377 | 85% | 1.1% | 0.0% | 0.4% | 1.5% | 13,250 |
| New York | Verizon- BA North | 12,050,789 | 89% | 3.0% | 14.5% | 2.2% | 19.6% | 2,365,206 |
| North Carolina | BellSouth | 2,603,650 | 50% | 2.3% | 1.2% | 1.4% | 4.9% | 128,832 |
| North Dakota | Qwest | 218,651 | 61% | 4.8% | 11.1% | 1.5% | 17.4% | 38,038 |
| Ohio | SBC- Ameritech | 4,063,464 | 59% | 1.3% | 0.7% | 2.4% | 4.5% | 181,924 |
| Oklahoma | SBC- SWBT | 1,660,815 | 83% | 3.2% | 1.5% | 0.3% | 4.9% | 81,690 |
| Oregon | Qwest | 1,460,169 | 66% | 1.7% | 2.6% | 2.2% | 6.5% | 94,355 |
| Pennsylvania | Verizon- BA South | 6,366,128 | 77% | 2.0% | 3.7% | 2.4% | 8.1% | 516,057 |
| Rhode Island | Verizon- BA North | 670,464 | 100% | 4.8% | 0.3% | 3.4% | 8.5% | 57,113 |
| South Carolina | BellSouth | 1,543,218 | 65% | 3.3% | 1.2% | 0.7% | 5.1% | 79,358 |
| South Dakota | Qwest | 280,799 | 65% | 3.9% | 6.2% | 0.4% | 10.6% | 29,676 |
| Tennessee | BellSouth | 2,764,068 | 80% | 1.8% | 1.2% | 1.6% | 4.6% | 125,973 |
| Texas | SBC- SWBT | 8,947,790 | 77% | 3.3% | 13.3% | 1.1% | 17.7% | 1,586,888 |
| Utah | Qwest | 1,165,099 | 95% | 1.9% | 0.2% | 0.8% | 2.8% | 33,070 |
| Vermont | Verizon- BA North | 368,392 | 85% | 4.9% | 0.1% | 0.1% | 5.1% | 18,898 |
| Virginia | Verizon- BA South | 3,800,149 | 76% | 3.3% | 0.0% | 3.1% | 6.5% | 245,516 |
| Washington | Qwest | 2,607,757 | 67% | 1.4% | 1.2% | 1.1% | 3.8% | 98,557 |
| West Virginia | Verizon- BA South | 897,968 | 84% | 1.4% | 0.0% | 1.0% | 2.4% | 21,407 |
| Wisconsin | SBC- Ameritech | 2,160,922 | 63% | 5.4% | 0.0% | 5.0% | 10.4% | 223,866 |
| Wyoming | Qwest | 261,266 | 83% | 1.0% | 8.2% | 0.0% | 9.1% | 23,900 |
| Nationwide BOC | | 141,311,809 | 75% | 3.1% | 3.0% | 1.6% | 7.6% | 10,765,430 |

*Source: FCC's ARMIS database. BOC Performance Metric Reports*
*Note: For some states CLEC purchased lines data is for June 1, not June 30.*

**Table 3: Percentage of ILEC Lines Purchased by CLECs, by Type of Line and Type of ILEC**

|  | Total ILEC Switched Access Lines | % of US Switched Access Lines | CLEC Lines Purchased as Percent of ILEC Lines | | | | Total CLEC Lines Purchased |
|---|---|---|---|---|---|---|---|
|  |  |  | Resale | UNE-P | UNE-L | Total |  |
| **December 2000** |  |  |  |  |  |  |  |
| Total ILEC | 186,501,328 | 100% | 2.7% | 1.7% | 1.1% | 5.5% | 10,243,113 |
| In-region BOCs | 141,311,809 | 75.8% | 3.0% | 2.2% | 1.2% | 6.4% | 9,110,970 |
|  |  |  |  |  |  |  |  |
| BOCs in states with 271 | 35,109,779 | 18.8% | 4.1% | 7.7% | 1.3% | 13.1% | 4,598,121 |
| BOCs in states w/o 271 | 106,202,030 | 56.9% | 2.6% | 0.4% | 1.2% | 4.2% | 4,512,849 |
| NonBOCs | 44,970,534 | 24.1% | 1.5% | 0.0% | 0.5% | 2.0% | 913,158 |
|  |  |  |  |  |  |  |  |
| BellSouth | 25,087,026 | 13.5% | 3.0% | 0.9% | 1.1% | 5.0% | 1,264,846 |
| Qwest | 17,626,160 | 9.5% | 1.3% | 2.0% | 0.7% | 3.9% | 696,126 |
| SBC- SWBT | 15,652,660 | 8.4% | 4.2% | 6.3% | 0.7% | 11.1% | 1,743,935 |
| SBC- Ameritech | 20,759,007 | 11.1% | 3.2% | 0.1% | 2.1% | 5.4% | 1,126,259 |
| SBC- PacTell | 19,191,553 | 10.3% | 1.2% | 0.0% | 1.3% | 2.5% | 483,610 |
| Verizon- BA North | 19,350,335 | 10.4% | 5.1% | 8.4% | 1.6% | 15.1% | 2,912,404 |
| Verizon- BA South | 23,645,068 | 12.7% | 1.4% | 1.2% | 1.1% | 3.7% | 883,790 |
|  |  |  |  |  |  |  |  |
| Verizon- GTE | 20,020,554 | 10.7% | 1.4% | 0.0% | 0.4% | 1.9% | 377,387 |
| SBC- SNET | 2,449,914 | 1.3% | 2.9% | 0.0% | 0.2% | 3.1% | 75,708 |
|  |  |  |  |  |  |  |  |
| **June 2001** |  |  |  |  |  |  |  |
| Total ILEC | 186,501,328 | 100% | N/A | N/A | N/A | N/A | N/A |
| In-region BOCs | 141,311,809 | 75.8% | 3.0% | 3.0% | 1.6% | 7.6% | 10,695,965 |
|  |  |  |  |  |  |  |  |
| BOCs in states with 271 | 35,109,779 | 18.8% | 3.4% | 9.3% | 1.7% | 14.4% | 5,067,723 |
| BOCs in states w/o 271 | 106,202,030 | 56.9% | 2.9% | 0.8% | 1.6% | 5.3% | 5,628,242 |
| NonBOCs | 44,970,534 | 24.1% | N/A | N/A | N/A | N/A | N/A |
|  |  |  |  |  |  |  |  |
| BellSouth | 25,087,026 | 13.5% | 2.8% | 1.6% | 1.2% | 5.6% | 1,393,714 |
| Qwest | 17,626,160 | 9.5% | 2.0% | 2.3% | 1.0% | 5.3% | 928,868 |
| SBC- SWBT | 15,652,660 | 8.4% | 3.7% | 8.5% | 0.8% | 13.0% | 2,035,343 |
| SBC- Ameritech | 20,759,007 | 11.1% | 2.8% | 1.6% | 2.8% | 7.2% | 1,488,741 |
| SBC- PacTell | 19,191,553 | 10.3% | 1.0% | 0.3% | 1.6% | 3.0% | 566,676 |
| Verizon- BA North | 19,350,335 | 10.4% | 4.0% | 9.2% | 2.1% | 15.2% | 2,944,727 |
| Verizon- BA South | 23,645,068 | 12.7% | 3.3% | 1.0% | 1.6% | 6.0% | 1,407,361 |
|  |  |  |  |  |  |  |  |
| Verizon- GTE | 20,020,554 | 10.7% | 1.2% | 0.0% | 0.5% | 1.7% | 348,539 |
| SBC- SNET | 2,449,914 | 1.3% | 3.1% | 0.0% | 0.6% | 3.8% | 92,477 |

*Source: ARMIS, BOC Performance Metric Reports, FCC 2001a*
*Note: For some states CLEC purchased lines data is for the beginning of the month.*
*Total BOC lines assumed to be the same for December and June.*
*N/A = Not Available*

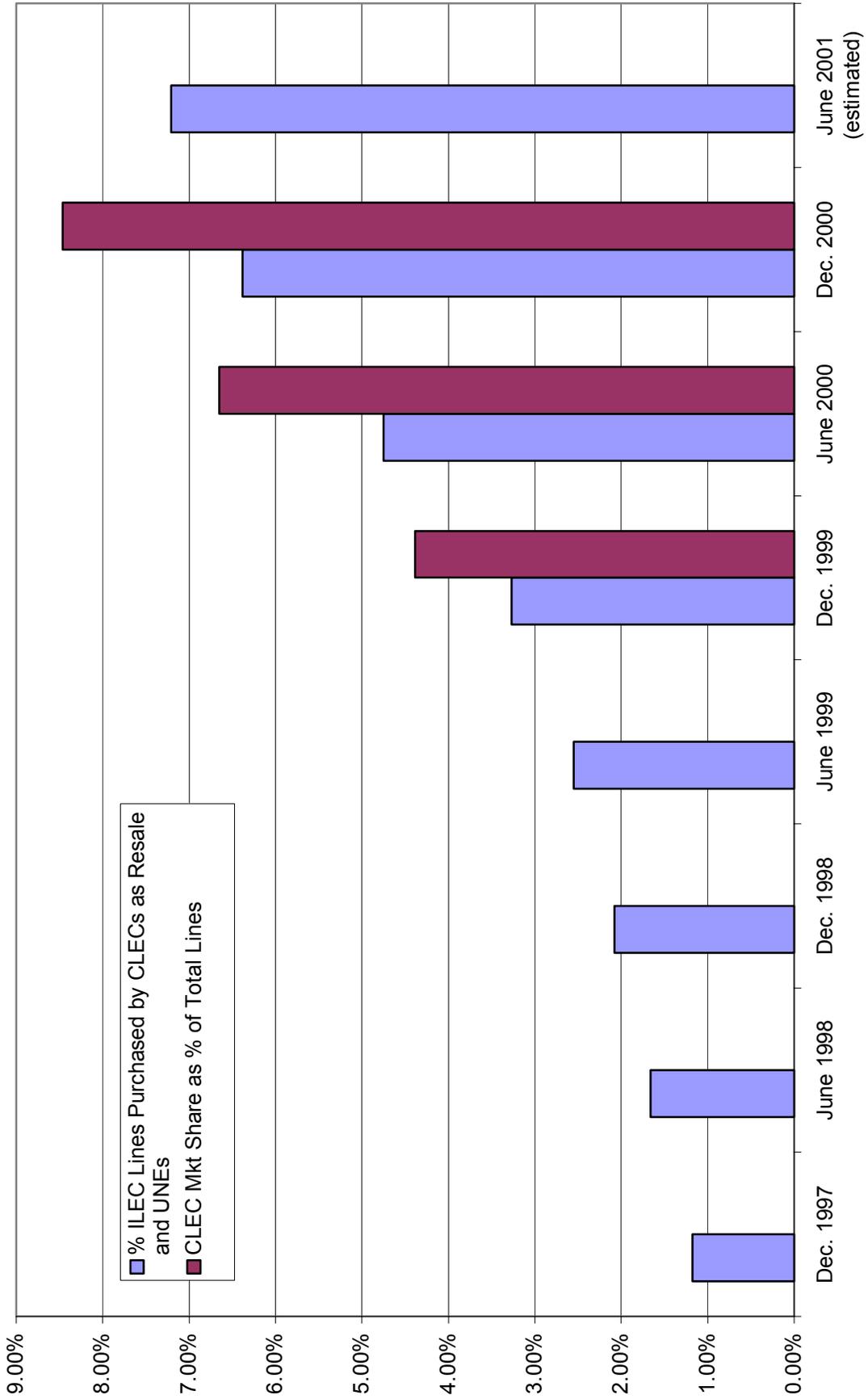

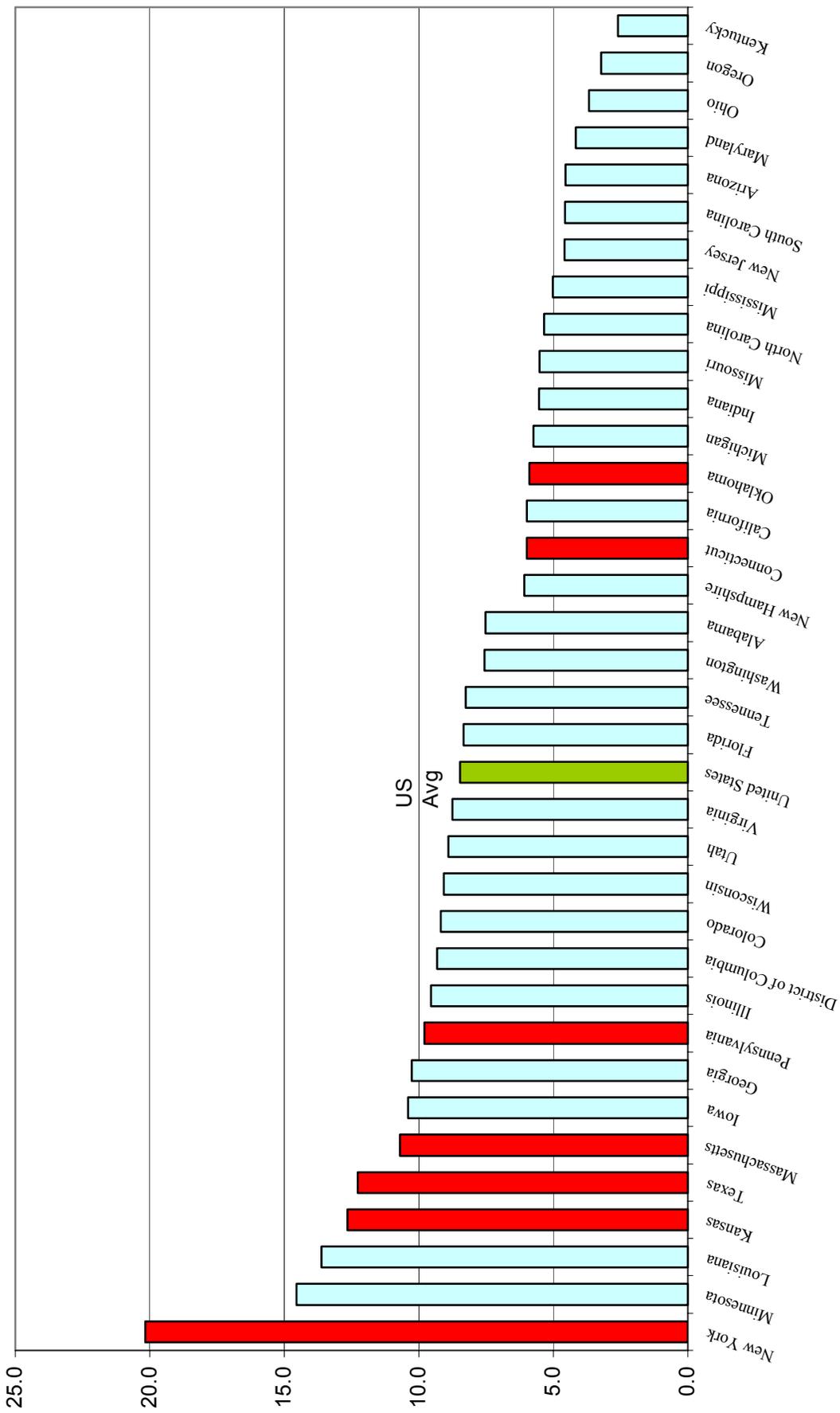

Figure 2: CLEC Market Share as of Dec. 2000 for States with Public Data
(States with granted 271 approval shaded darker)

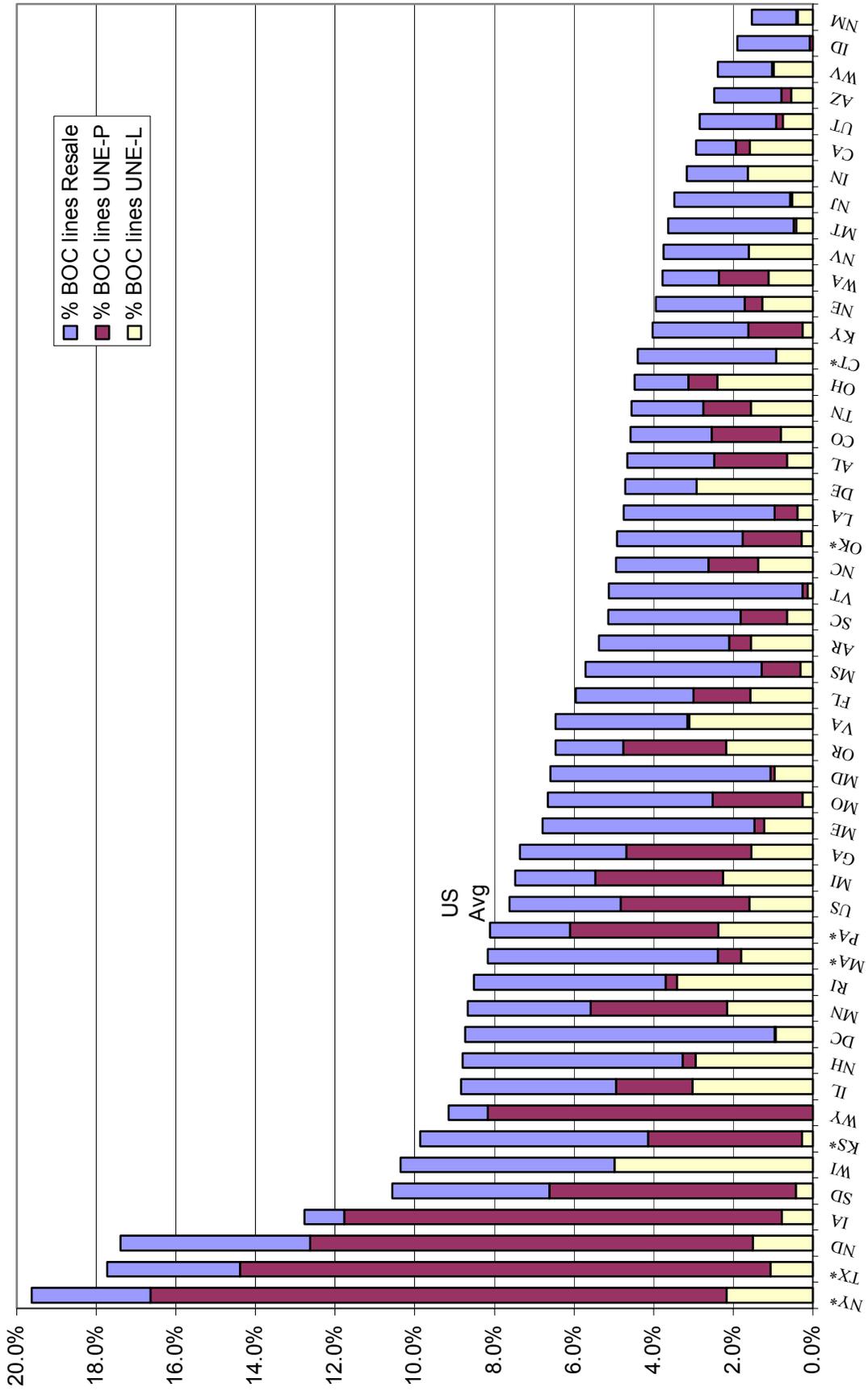

Figure 3: BOC Lines purchased by CLECs for POTS as Percent of BOC Lines, June 2001 (States with granted 271 approval marked with *)

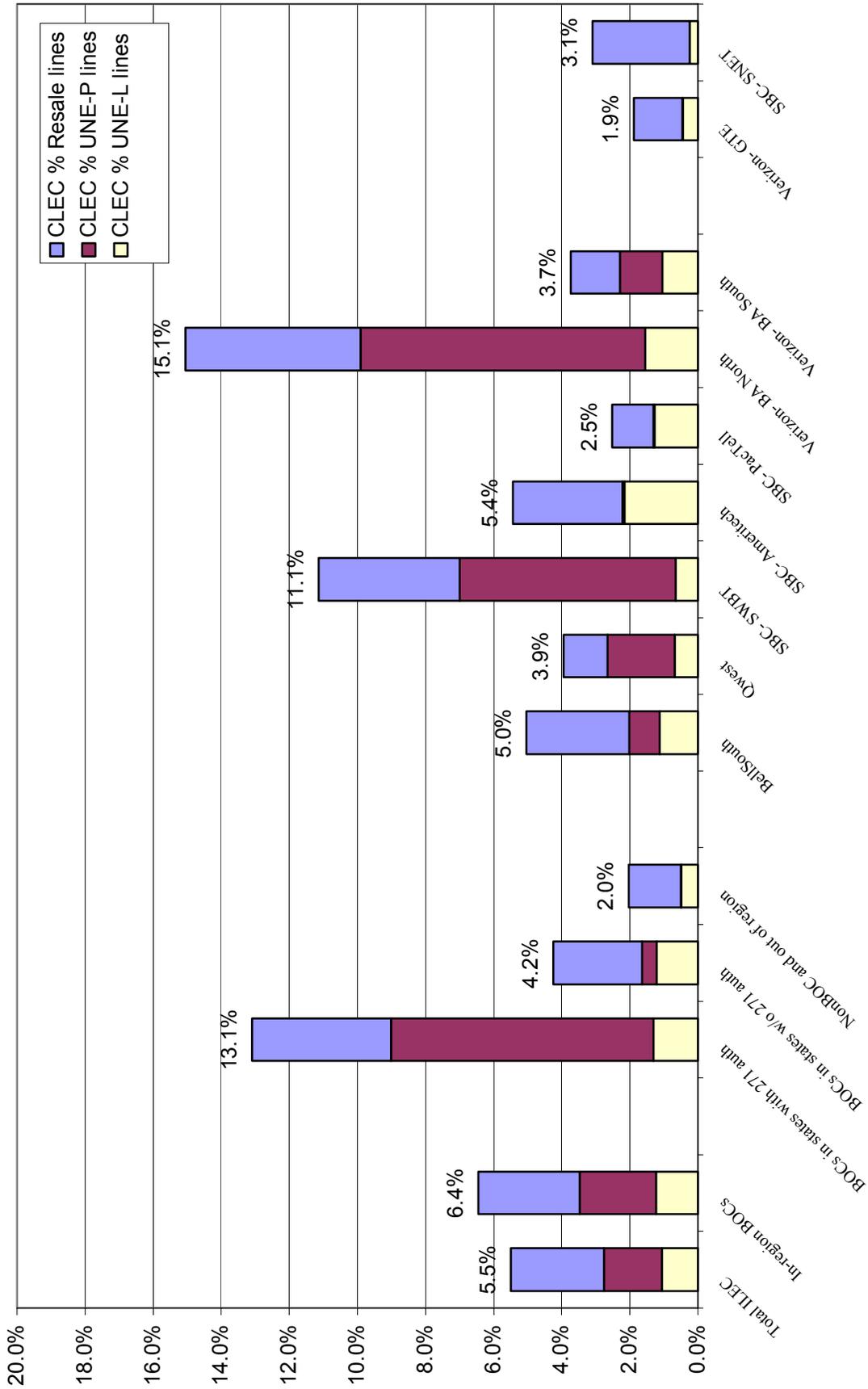

Figure 4: ILEC Lines Purchased by CLECs for POTS as Percent of ILEC Lines, Dec. 2000